\documentclass[
    ,final            
  ]
  {aipproc}

\usepackage{color}

\layoutstyle{8x11single}

\begin{document}

\title{Thermal diffusion segregation of an impurity in a driven granular fluid}


\author{Francisco Vega Reyes}{
  address={Departamento de F\'isica, Universidad de Extremadura, E-06071 Badajoz, Spain},
altaddress={Instituto 
de Computaci\'on Cient\'ifica Avanzada 
(ICCAEx), Universidad de Extremadura, E-06071 Badajoz, Spain}
}

\author{Vicente Garz\'o}{
  address={Departamento de F\'isica, Universidad de Extremadura, E-06071 Badajoz, Spain},
altaddress={Instituto 
de Computaci\'on Cient\'ifica Avanzada 
(ICCAEx), Universidad de Extremadura, E-06071 Badajoz, Spain} 
}

\begin{abstract}
We study segregation of an impurity in a driven granular fluid under two types of \emph{steady} states. In the first state, the granular gas is driven by a stochastic volume force field with a Fourier-type profile
while in the second state, the granular gas is sheared in such a way that inelastic cooling is balanced by viscous heating. We compare theoretical results derived from a solution of the (inelastic) Boltzmann equation at Navier-Stokes (NS) order with those obtained from the Direct Monte Carlo simulation (DSMC) method and molecular dynamics (MD) simulations. Good agreement is found between theory and simulation, which provides strong evidence of the reliability of NS granular hydrodynamics for these steady states (including the dynamics of the impurity), even at high inelasticities. In addition, preliminary results for thermal diffusion in granular fluids at moderate densitis are also presented. As for dilute gases \cite{VGK14}, excellent agreement is also found in this more general case.

\end{abstract}

\maketitle


\section{Introduction}
\label{intro}

Granular gas dynamics is a subject of interest in a number of fundamental research fields and industry applications. For this reason, a large number of works have dealt with this kind of systems \cite{G03}. By granular gas we mean  a set of mesoscopic particles at low density that lose a fraction of their kinetic energy after collisions with other particles. The number density is low enough so that collisions are always binary and instantaneous (compared to the characteristic time between collisions) and thus, the system obeys the Boltzmann (or Enskog) kinetic equation, with the appropriate modification that takes into account inelasticity in collisions. For this reason, this kind of system is usually called granular ``gas''. In this work we will consider two distinct sets of inelastic spheres (granular binary mixture), one of them being present in a much smaller concentration than the other one (tracer limit). This is equivalent to study the dynamics of a ``granular impurity'' (of mass $m_0$ and diameter $\sigma_0$) in a granular gas constituted by particles of mass $m$ and diameter $\sigma$. The granular impurity would obey the Boltzmann-Lorentz kinetic equation (since collisions among impurity particles themselves can be neglected) while  the granular gas would obey the (closed) Boltzmann kinetic equation (since the state of the gas is not affected by the presence of impurities) \cite{VGS08}.

We are here mainly interested in studying transport properties induced in the system by the presence of a temperature gradient. To achieve a \emph{steady} state, two different types of energy inputs will be considered. In the first case (Case I), particles of the system are assumed to be heated by a stochastic driving force that exactly compensates for the inelastic cooling produced by collisions. This kind of external force attempts to model the interaction of the system (impurity plus granular gas) with a surrounding fluid that subjected to this configuration shows a Fourier temperature profile \cite{VGK14}. In the second case (Case II), the gas is sheared from the boundaries so that, viscous heating is balanced by the inelastic cooling. As said before, we will describe in Case I the thermal effects of the interstitial fluid on the granular particles by means of a stochastic volume force (in our case, a white noise) that will act more strongly in the hotter areas of the surrounding fluid and more weakly in its colder areas. For this reason, we will assume that the volume force has a Fourier-like profile. This type of forcing has been shown to exactly balance the effects of collisional cooling, inducing a steady state with uniform heat flux \cite{VSG10}. We will show that the steady state hydrodynamic profiles as well as the transport coefficients are correctly described by NS hydrodynamic theory if the appropriate Chapman-Enskog method is followed to solve the kinetic equation to first order in spatial gradients. In this case, we use a procedure where exact balance between inelastic cooling and thermostat heating is achieved locally in the zeroth-order balance equations \cite{GM02}.

We present in this work additional results for the temperature ratio and thermal diffusion segregation not previously reported in Ref.\ \cite{VGK14}. In particular, we compare theoretical results derived from the Boltzmann (low-density limit) and Enskog (moderate densities) kinetic equations with those obtained from the Direct Simulation Monte Carlo (DSMC) method \cite{B94}.
Good agreement is found for a dilute system and, interestingly, also in preliminary results for a granular fluid at moderate densities.

\section{Description of the system and steady base states}
\label{base}

Our system consists of a large number of identical hard disks/spheres ($d=2,~d=3$, respectively). The particles collide loosing a fraction of their kinetic energy; i.e., collisions are inelastic. The degree of inelasticity is accounted for by the (constant) coefficient of normal restitution $\alpha \leq 1$. We assume also that the particles are distributed in space with low density at all times. As we said, depending on the value of the density, the granular gas may be described by the inelastic Boltzmann kinetic equation (which only strictly applies for vanishing density) and/or the Enskog kinetic equation (which applies for moderate densities) \cite{FK72}. We mix this set of granular particles with another set composed also of identical granular particles but in general mechanically different to the particles of the first set and additionally, with a negligible concentration. For this reason, will refer to this second set as 'granular impurities'. Collisions between impurity and gas particles are also inelastic and are characterized by another (constant) coefficient of restitution, $\alpha_0\le1$. Given that the relative concentration of impurities is much smaller than that of gas particles (and so, impurity-impurity collision events are not statistically relevant compared to gas-impurity collisions), one can assume that the one-particle velocity distribution function of impurities verifies the (linear) Boltzmann-Lorentz kinetic equation \cite{VGS08}. In addition, the collisions between impurity and gas particles can be also neglected in the kinetic equation of granular gas and hence, its one-particle velocity distribution function obeys the nonlinear (closed) Boltzmann equation.

We assume that both sets of particles are enclosed between two infinite parallel walls that act as distinct temperature sources, thus heating the system and yielding a temperature gradient from the system boundaries. The walls are located at planes $y=\pm h/2$ respectively. We denote the only relevant parallel direction to the walls as $x$. For explicit expressions of the kinetic equations that apply for our system please refer to a previous work where the same system is analyzed \cite{VGK14}. We consider also that energy input is strong enough so as to make negligible the effects of gravity on the granular particles.

By taking moments in the Boltzmann kinetic equation for the granular gas we obtain the relevant balance equations \cite{VGK14}. For our geometry, the energy balance equation takes the form
\begin{equation}
\label{qbal}
\frac{1}{\nu}\frac{\partial q_y}{\partial y}=-\frac{d}{2\nu}nT(\zeta-\sigma_T)-P_{xy} \frac{1}{\nu}\frac{\partial U_x}{\partial y} \equiv \mathcal{C},
\end{equation}
where$\nu\equiv p/\eta_0$ (with $\eta_0=\sqrt{mT}(d+2)\Gamma (d/2)\pi^{-(d-1)/2}/8$) is an effective collision frequency of granular gas and $\mathcal{C}$ is a constant \cite{VSG13}. Also, $q_y$ is the $y$ component of the the heat flux, $n$ is the number density of gas particles, $T$ is the granular temperature, $\zeta$ is the inelastic cooling rate, $\sigma_T$ is an energy source, $P_{xy}$ is the $xy$ element of the  stress tensor and $U_x$ is the flow velocity in the horizontal direction. For this work we will only consider steady states with \emph{uniform} heat flux, and thus $\mathcal{C}=0$ \cite{VGK14}. As we already mentioned, two different choices for $\sigma_T$ will be considered in this work:

(a) \textit{\underline{Case I:}} $U_\pm=0$ (no shear, and thus in Eq.\ \eqref{qbal} the viscous heating term $P_{xy}\partial_x U_y$  vanishes). In this case, the inelastic cooling in the system is compensated by the heat injected by the external force. Here, we assume that the system is driven by means of a stochastic Langevin force representing a Gaussian white noise \cite{WM96}. The term associated with the external force is written in the corresponding kinetic equation as a Fokker-Planck operator \cite{NE98} of the form ${\cal F}\equiv -\frac{1}{2}\xi^2\partial^2/\partial v^2$. The covariance of the stochastic acceleration $\xi^2$ is chosen to be the same for both species (impurity and gas particles) \cite{GV12}. Thus, in Case I, the production of energy term is  $\sigma_T=m\xi^2/T$. The stochastic external forcing is frequently used in computer simulations \cite{PLMPV98,CLH00,KSSAOB05,CBMNS06,FAZ09,KSZ10} and has been also proved experimentally \cite{KSSAOB05,OLDLD04}.

(b) \textit{\underline{Case II}}:  No external driving ($\sigma_T=0$) but $U_-\neq U_+\neq0$, i.e., when both walls are in relative motion (sheared granular gas). In this case, inelastic cooling is compensated by viscous heating. Surprising as it may be, a local balance between inelastic cooling and viscous heating may be globally achieved \cite{VSG10}.

Both cases satisfy the following differential equations for the relevant hydrodynamic profiles \cite{VGK14, VSG13}:
\begin{eqnarray}
& &\frac{\partial u_{y}}{\partial y}=0,
\label{profu}\\
& &\frac{1}{\nu}\frac{ \partial}{\partial y}\left(\frac{1}{\nu}\frac{\partial  T}{\partial y}\right)=0.
\label{profT}
\end{eqnarray}

Respect to the state of the impurity, and as we said before, the thermal diffusion factor $\Lambda$ characterizes the amount of segregation parallel to the thermal gradient. It is thus defined by
\begin{equation}
\Lambda=-\frac{\left(\frac{\partial \ln (n_0/n)}{\partial y}\right)}{\left(\frac{\partial \ln T}{\partial y}\right)}.
\label{segcrit}
\end{equation}
The thermal diffusion coefficient $\Lambda$ depends on the mechanical properties of the granular gas and impurity and the temperature ratio $\chi=T_0/T$, where $T_0$ is the granular impurity temperature \cite{VGK14}. This dependence is expressed in the form: $\Lambda\equiv \Lambda(\Xi,\chi)$ with $\Xi\equiv \left\{\alpha,\alpha_0,\mu,\omega\right\}$. Here, $\mu\equiv m_0/m$ and  $\omega\equiv \sigma_0/\sigma$ are the mass and diameter of the impurity particles relative to those of the granular gas particles, respectively. As in our previous work, and since $\Lambda$ is strongly dependent on the temperature ratio value, we make the following ansatz: for both Cases I and II and to get $\Lambda$, we use the tracer diffusion transport coefficients obtained from the Chapman-Enskog method at first order in the gradients (i.e., NS order) for the granular system heated by the stochastic force \cite{GV12}. In addition, we insert different forms of $\chi$ for both cases. In Case I, $\chi_I$ is obtained from the condition $\chi_I m \zeta_0 =m \zeta$, where $\zeta_0$ is the cooling rate associated with the impurity temperature $T_0$ \cite{GV12}. In case II, $\chi_{II}$ is obtained from Grad's moment method (non-linear theory) for sheared granular gases with uniform heat flux \cite{G02}.


\section{Results and discussion}

\begin{figure}
\centering
\includegraphics[height=4.5cm]{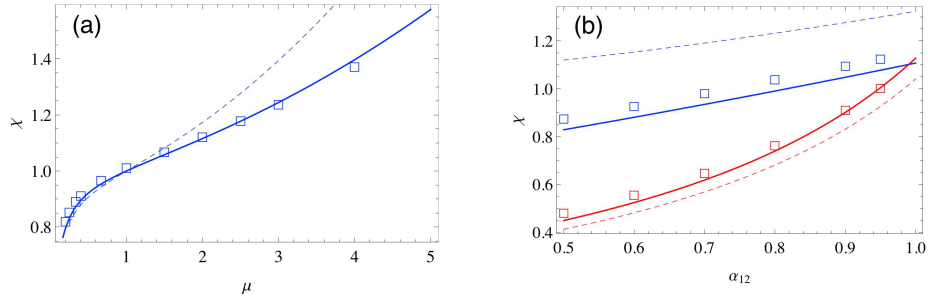}
\caption{Temperature Ratio $\chi=T_0/T$ for sheared spheres (case II). In panel (\textit{a}) we represent $\chi$ vs. relative mass $\mu$, for the case $\alpha=\alpha_0=0.9$, $\omega=1$. In panel (\textit{b}) we represent $\chi$ vs. $\alpha_0$ with $\alpha=0.9$ for two cases: $\mu=4$, $\omega=2$ (blue) and $\mu=1/4$, $\omega=1/2$ (red). In both panels, symbols stand for DSMC data, dashed lines stand for NS theory and solid lines for Grad's theory results. It can be noticed that the temperature ratio is better predicted by the non-linear Grad's theory.}
\label{fig2}\end{figure}

In this Section, we will compare theoretical results with those obtained from the Monte Carlo (DSMC) method and molecular dynamics (MD) for cases I and II. First, we consider the dependence of the temperature ratio $\chi\equiv T_0/T$ on collisional dissipation. The quantity $\chi$ measures the breakdown of energy equipartition in a granular mixture. In Figure \ref{fig2} we consider a sheared gas (Case II) with different values of the mass $\mu$ and diameter $\omega$ ratios and coefficients of restitution. For the sake of completeness, we have plotted the theoretical results obtained for Cases I ($\chi_I)$ and II ($\chi_{II})$. As we can see, only for $\chi_{II}$ we obtain a good agreement with simulations. Furthermore, this agreement is excellent in most cases, as it can be noticed. On the other hand, the comparison with $\chi_{I}$ is never as good, which justifies our ansatz. In panel (\textit{a}) the temperature ratio $\chi$ is plotted vs. relative mass $\mu\equiv m_0/m$ for equal coefficients of restitution. The agreement is always very good, although it is a little better for lighter impurities ($\mu<1$) and the theory line tends to separate for heavy impurities ($\mu\ge 4$). The same can be deduced from results in panel (\textit{b}), where we have represented a case of varying $\alpha_0$ for constant $\alpha$, in order to check the agreement for cases with $\alpha\neq\alpha_0$. The results show also an excellent agreement for the whole range of values of $\alpha_0$ plotted here (that include strong inelasticity values, like $\alpha=0.5$). Two cases are represented here $\mu=4, \omega=2$ (in blue) and $\mu=1/4, \omega=1/2$ (in red). As in panel (\textit{a}), the results tend to be slightly better for lighter impurities (in red). In any case the agreement is overall very good again.

\begin{figure}
\centering
\includegraphics[height=4.5cm]{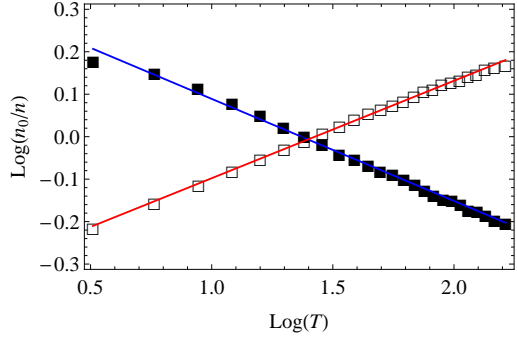}
\caption{Check of thermal diffusion factor, from DSMC simulations ($\ln {n_0/n}$ vs. $\ln T$. In this case, we represent results for spheres with $\alpha=\alpha_0=0.9$ and $\omega=1$, for $\mu=2$ (solid symbols) and $\mu=1/2$ (open symbols). In both cases we obtain the numerical value of the thermal diffusion factor $\Lambda$ from a linear fit, as predicted by hydrodynamic theory. It can be noticed that this linear fit is excellent, which implies that $\Lambda$ is in effect constant throughout the system.}
\label{fig3}\end{figure}

Next, we consider the thermal diffusion factor $\Lambda$. First, we want to check whether this quantity is \emph{uniform} in the bulk domain of the system. If this is the case, our segregation criterion will be a global feature of the bulk region and so, it is not restricted to specific regions of the system. In general, our simulations show that actually  $\Lambda$ is constant. As an illustration, the profile of $\ln {n_0/n}$ vs. $\ln T$ is plotted in Figure \ref{fig3} for a sheared mixture (Case II). According to the theory (see Eq. \ref{segcrit}), the profile $\ln T$ should be linear when represented vs. $\ln {n_0/n}$. It is apparent that this prediction is fulfilled with a high degree of accuracy, even in the regions near the boundaries (i.e., boundary layer effects are not significant regarding with granular impurity segregation). Two cases have been represented here, both for $\omega=1$, $\alpha=\alpha_0=0.9$: $\mu=2$ (solid symbols) and $\mu=1/2$ (open symbols). Results also confirm the general trend observed in a previous work \cite{VGK14}, where we observed that except for smaller impurities ($\omega\le 2/3$) the granular heavier impurities tend to segregate towards the colder wall, which would correspond to a negative slope in a $\ln T$ vs $\ln {n_0/n}$ for the $\mu=2$ series and a positive slope for the $\mu=1/2$ series, as obtained in Figure \ref{fig3}.

\begin{figure}
\centering
\includegraphics[height=4.5cm]{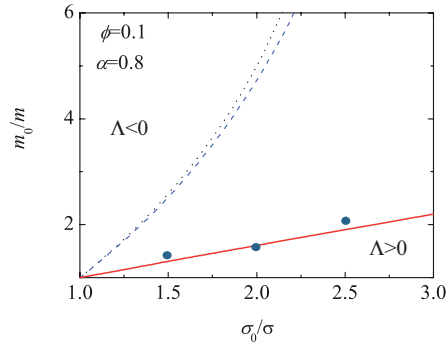}
\caption{Plot of the marginal segregation curve ($\Lambda=0$) for a granular impurity in a heated granular gas (Case I) with finite density  (volume fraction  $\phi=0.1$) for $\alpha=\alpha_0=0.8$. Lines stand for theory results and symbols for DSMC data. The blue lines stand for first Sonine approach (dotted line for standard Sonine approximation and dashed line for modified Sonine, both to first order). The red line stands for second Sonine approach. As we can see, second order terms in the Sonine polynomial expansion are needed in order to get a good agreement with simulations that, as we can see, at this level of approximation is excellent. }
\label{fig4}\end{figure}

Now we consider segregation in a moderately dense granular gas driven by the stochastic thermostat (Case I). In a previous work \cite{VGK14}, we have shown that the agreement between theory and simulation for the thermal diffusion factor $\Lambda$ is excellent for both types of flows (case I and II) in the low-density regime. Figure \ref{fig4} shows the marginal segregation curve ($\Lambda=0$) for a granular gas outside the dilute limit (more concretely, we present a gas where the particles occupy $10\%$ of the system volume; i.e. the volume fraction is $\phi=0.1$). The theoretical results for this figure have been extracted from the expression of $\Lambda$ derived in reference \cite{GV12}. We represent two different levels of approach in the Sonine polynomial expansion used in the Chapman-Enskog method. As we can see, the agreement with simulations at the level of the second order Sonine approximation is excellent. To our knowledge, this is the first evidence of NS theory accuracy for a granular segregation problem outside the dilute limit and for moderate inelasticity ($\alpha=0.8$ in this case)

\begin{figure}
\centering
\includegraphics[height=4.5cm]{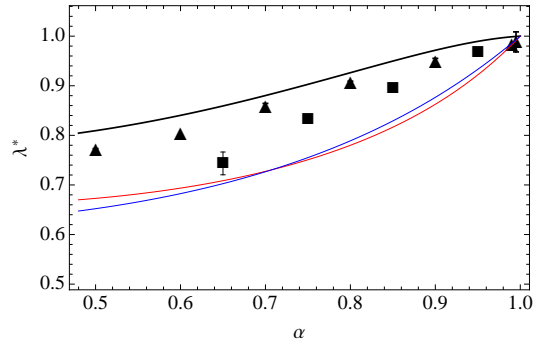}
\caption{Plot of the reduced thermal conductivity coefficient ($\lambda^*$) for a granular impurity in a sheared granular gas in the dilute limit  (null volume fraction  $\phi=0.0$). The transport coefficient is represented vs. coefficient of restitution $\alpha$ for a wide range of inelasticities; from the elastic limit ($\alpha=1$) to very strong inelasticities ($\alpha=0.5$). Symbols stand for simulations: triangles denote DSMC data and squares denote molecular dynamics simulations (both for case I). Lines stand for theory results. The black line stands for the non-linear theory results (Grad's moments method, from reference \cite{VSG10}) and the colored lines stand for NS results for case I steady states.}
\label{fig5}\end{figure}

Apart from the fact that both types of steady states have uniform heat flux, one might be surprised that the NS segregation criteria show such an excellent performance, specially for high inelasticities, for which in general is expected that hydrodynamics at NS order would fail. This failure of NS hydrodynamics for steady granular flows has been discussed in a number of works in the field \cite{G03}. Our guess is that since for both cases the impurity segregation is originated by a temperature gradient, and both cases I and II have identical temperature profiles, one could expect a good agreement if the response of the system to this identical temperature gradient is also similar. One could measure the response of the system to the temperature gradient by analyzing the behavior of the thermal conductivity coefficient $\lambda$. At the level of NS hydrodynamics and for the geometry in our system, the thermal conductivity is defined by the relation $q_y=-\lambda\partial T/\partial y$. In Figure \ref{fig5} we plot the reduced thermal conductivity $\lambda^*=\lambda/\lambda_0$ where $\lambda_0$ is the thermal conductivity for the elastic gas \cite{CC70}. More concretely we compare the results for: 1) Grad's theory for case II flows, 2) NS theory for case I flows, 3) simulations (both molecular dynamics and DSMC) for case II flows. It is interesting to note that the theoretical curves corresponding to Grad's theory for sheared states from case II and NS theory for non-sheared states from case I present similar qualitative behavior with respect to inelasticity since both are monotonically decreasing for increasing inelasticity (decreasing $\alpha$). This seems to indicate that, as we hypothesized, the response of the granular gas to the temperature gradient is qualitatively similar for both types of states. This could cast some light in the search an explanation on the similarity of the impurity segregation behavior that we have found in the present and a previous work \cite{VGK14}. This similarity is specially surprising since we have detected no similarity in the values of the respective collision frequencies associated to heat flux for both theories \cite{VSG10,GM02}.

In spite of this remaining lack of explanation, the additional results presented in this work confirm that NS hydrodynamics correctly describes important transport properties of steady granular flows with uniform heat flux. This agreement is fulfilled almost independently of the degree of inelasticity, contrary to what has been argued in a large number of previous works on granular dynamics \cite{G03}. We plan to extend the study on impurity segregation in granular gases at moderate densities in a forthcoming work (extending the results of Figure \ref{fig4}). We also expect to analyze to what extent NS hydrodynamics can also be applied for highly inelastic granular gases for other systems.

\begin{theacknowledgments}
F. V. R. and V. G. acknowledge support of the Spanish Government through Grants FIS2010-12587 and  FIS2013-42840 and the Junta de Extremadura through Grant No. GR10158, partially financed by FEDER funds. The authors also thank Nagi Khalil for fruitful discussion.
\end{theacknowledgments}

\bibliographystyle{aipproc}
\bibliography{FVR_VG}

\end{document}